\documentclass[aps,prb,twocolumn,10pt,superscriptaddress,floatfix,longbibliography]{revtex4-2}

\usepackage{amsmath,amssymb} 
\usepackage{bm} 
\usepackage{graphicx} 
\usepackage{comment} 
\usepackage{textcomp} 
\usepackage{slashed}

\usepackage{enumitem}
\setlist{noitemsep,leftmargin=*,topsep=0pt,parsep=0pt}

\usepackage{xcolor} 
\definecolor{lightgray}{gray}{0.6}
\definecolor{medgray}{gray}{0.4}

\usepackage{hyperref}
\hypersetup{
colorlinks=true,
urlcolor= blue,
citecolor=blue,
linkcolor= blue,
}

\newif\ifptitle
\newif\ifpnumber
\newcounter{para}

\ptitletrue  
\pnumbertrue  

\newcommand{\mytitle}{A New Form of Soft Supersymmetry Breaking?}

\begin{document}

\title{\mytitle}

\author{Scott Chapman}
\email[]{schapman@chapman.edu}
\affiliation{Institute for Quantum Studies, Chapman University, Orange, CA  92866, USA}

\date{\today}

\begin{abstract}
Starting with a supersymmetric U(N)xU(N) gauge theory built in N=1 superspace, a nonsupersymmetric theory is obtained by ``twisting'' the gauginos into a different representation of the group than the gauge bosons. Despite the fact that this twisting breaks supersymmetry, it is still possible to construct an action that is holomorphic and invariant to local ``twisted'' gauge transformations in superspace.  It is conjectured that these two properties may allow the theory to be free of quadratic divergences to all orders, despite a lack of supersymmetry.  An explicit calculation shows that the theory is free of quadratic divergences to at least the two-loop order.      
\end{abstract}

\maketitle

One of the motivations for supersymmetry (SUSY) is the fact that supersymmetric theories are free of quadratic divergences.  As such, they provide a resolution (at least partially) to the Hierarchy problem.  But a theory does not need to be completely supersymmetric to be free of quadratic divergences.  For many years it has been known that certain types of ``soft'' supersymmetry breaking terms may be added to a supersymmetric action without introducing quadratic divergences \cite{Grisaru-soft,Martin-soft}.  These SUSY-breaking terms allow phenomenological supersymmetry models to reproduce the observed particle spectrum.  For example, soft SUSY-breaking terms are incorporated into the Minimal Supersymmetric Standard Model \cite{MSSM-1986,MSSM-1985,MSSM-2008}. 
 
Even though they break supersymmetry, the soft terms referenced above can be derived from actions with unbroken supersymmetry at a higher scale.  As a result, the soft terms do not introduce a violation of the theorem by Haag, Lopuszanski, and Sohnius (HLS) \cite{HLS} which states that superpartners (in N=1 supersymmetric actions) must be in the same representation of the gauge group.  That requirement puts a severe restriction on attempts to match the observed particle spectrum with boson-fermion superpartners in phenomenological models.  In fact, most believe that it rules out the possibility that any of the already-detected particles are superpartners with each other.  For example, the HLS requirement would prohibit the gauge bosons of the Standard Model from being superpartners with left-handed quarks, since the former are in the adjoint representation while the latter are in the (3,2) representation of the gauge group.

This paper explores a method of supersymmetry breaking that causes the ``superpartners'' to be in different representations of the gauge group.  The specific supersymmetry breaking terms (the changes made to a supersymmetric action to arrive at the action of this paper) are limited to modifications of gaugino interaction vertices and are shown in eqs. \eqref{22}, \eqref{23}, and \eqref{24} below.

Although the action considered in this paper is not supersymmetric, it is still invariant to local “twisted” superspace gauge transformations, and the pure gauge part of the action is holomorphic.  Some of the proofs of nonrenormalization theorems for supersymmetric theories rely heavily on gauge invariance and the pure gauge part of the action being holomorphic (see for example \cite{SQCD-Argyres}).  It is conjectured here that many of those arguments could be applied to the theory presented in this paper, and that consequently, the theory may be free of quadratic divergences.  A proof (or disproof) of this conjecture is outside the scope of this paper.  The more limited scope of this paper is to show that the theory is free of quadratic divergences to at least the two-loop order.

The theory under consideration is defined starting in eq \eqref{6} below.  To motivate the structure of the theory, one may begin by considering the real superfield for a supersymmetric gauge theory involving the group $U(N)_1\times U(N)_2$:
\begin{equation}\label{1}
\hat{V}=\left(\begin{array}{cc}{V_1} & {0} \\ {0} & {V_2}
\end{array}\right),
\end{equation}
where $V_1=V_1^A\left(x,\theta,\bar{\theta}\right)t^A$ is a real superfield in the adjoint representation of the group $U(N)_1$. Similarly, $V_2$ is a real superfield in the adjoint representation of the group $U(N)_2$. The group structure of these fields is defined by $t^A$, the $N\times N$ fundamental-representation matrices of $U(N)$, normalized by ${\rm{tr}}\left(t^At^B\right)=\tfrac{1}{2}\delta^{AB}$.  For reviews involving superfields, see \cite{rargurio,SUSY-Martin,SUSY-Haber,Binetruy,SQCD-Argyres}.  Supergauge transformations for this real superfield take the form:
\begin{equation}\label{2}
e^{2g\hat{V}}\to e^{i\hat{\Lambda}^{\dag}}e^{2g\hat{V}}e^{-i\hat{\Lambda}},
\end{equation}
where $\hat{\Lambda}$ has the same block-diagonal structure as $\hat{V}$, and   $\Lambda_m=\Lambda_m^A\left(y,\theta\right)t^A$ are chiral superfield functions of  $\theta_\alpha$ and $y^\mu = x^\mu +i\theta\sigma^\mu \bar{\theta}$.  Notational conventions of \cite{rargurio} are used throughout.  All of the calculations of this paper are assumed to take place at a unification scale where the same coupling constant $g$ can be used for all fields (both Abelian and non-Abelian).

The real superfield can be expanded in terms of its component fields as follows:
\begin{equation}\label{3}
\begin{aligned}
\hat{V}=&\hat{C}+\hat{\eta}\hat{\theta}+\hat{\bar{\theta}}\hat{\bar{\eta}} 
+\hat{N}\hat{\theta}^2+\hat{\bar{\theta}}^2\hat{N}^{\dag} \\
-&\hat{\bar{\theta}}\bar{\sigma}^\mu \hat{A}_\mu \hat{\theta}
+i\hat{\bar{\theta}}\hat{\bar{\lambda}}\hat{\theta}^2
-i\hat{\bar{\theta}}^2\hat{\lambda}\hat{\theta}
+\tfrac{1}{2}\hat{\bar{\theta}}^2\hat{d}\hat{\theta}^2 ,
\end{aligned}
\end{equation}
where each component field is a function of spacetime with the same block-diagonal matrix structure as $\hat{V}$.  For example, $\hat{\eta}$ has block diagonal components $\eta_1$ and $\eta_2$ where $\eta_m=\eta_m^A\left(x\right)t^A$.  Also, the following notation is utilized in eq \eqref{3}:
\begin{equation}\label{4}
\hat{\theta}_\alpha =
\left(\begin{array}{cc}
{\theta_\alpha} & {0} \\ {0} & {\theta_\alpha}
\end{array}\right)
\,\,\,\,\textrm{and}\,\,\,\,
\hat{\bar{\theta}}_{\dot{\alpha}}=
\left(\begin{array}{cc}
{\bar{\theta}_{\dot{\alpha}}} & {0} \\ {0} & {\bar{\theta}_{\dot{\alpha}}}
\end{array}\right)
\end{equation}
The chiral gauge superfield $\hat{\Lambda}$ can similarly be expanded in terms of  $\hat{\theta}_\alpha$.

One may consider ``twisting'' the Grassman coordinates in the following way:
\begin{equation}\label{5}
\hat{\theta}_\alpha\to
\left(\begin{array}{cc}
{0} & {\theta_\alpha} \\ {\theta_\alpha} & {0}
\end{array}\right)
\,\,\,\,\textrm{and}\,\,\,\,
\hat{\bar{\theta}}_{\dot{\alpha}}\to
\left(\begin{array}{cc}
{0} & {\bar{\theta}_{\dot{\alpha}}} \\ {\bar{\theta}_{\dot{\alpha}}} & {0} 
\end{array}\right)
\end{equation}
If this was used on the expansion of $\hat{V}$ in eq \eqref{3}, it would put all of the boson (fermion) component fields of $\hat{V}$ in the diagonal (off-diagonal) $N\times N$ blocks of $\hat{V}$.  Applying the same modification to $\hat{\Lambda}$ would give it the same boson/fermion matrix structure.  Since multiplying any two matrices with this structure results in another matrix with this structure, gauge transformations defined by eq \eqref{2} would maintain that structure.  Therefore, such gauge transformations would not necessarily be inconsistent, and it should be possible to construct actions invariant to them.  However, the modification of eq \eqref{5} breaks supersymmetry, so any action built using that modification would not be supersymmetric.  

With that motivation, it is possible to define a real ``twisted superfield'' (that is strictly speaking not actually a superfield) in the following way:
\begin{widetext}
\begin{equation} \label{6} 
V=\left(\begin{array}{cc} {C_{1} +N_{1} \theta ^{2} +\bar{\theta }^{2} N_{1}^{\dag } -\bar{\theta }\bar{\sigma }^{\mu } A_{1\mu } \theta +{\tfrac{1}{2}} \bar{\theta }^{2} d_{1} \theta ^{2} } 
& {\eta_1 \theta +\bar{\theta }\bar{\eta}_2 +i\bar{\theta }\bar{\lambda}_2 \theta ^{2} -i\bar{\theta }^{2} \lambda_1 \theta } \\ 
{\eta_2 \theta +\bar{\theta }\bar{\eta}_1 +i\bar{\theta }\bar{\lambda}_1 \theta ^{2} -i\bar{\theta }^{2} \lambda_2\theta } 
& {C_{2} +N_{2} \theta ^{2} +\bar{\theta }^{2} N_{2}^{\dag } -\bar{\theta }\bar{\sigma }^{\mu } A_{2\mu } \theta +{\tfrac{1}{2}} \bar{\theta }^{2} d_{2} \theta ^{2} } \end{array}\right). 
\end{equation} 
\end{widetext}
In the above $2\times 2$ matrix, each of the component fields is an $N\times N$ matrix function of $x$.

The ``twisting'' has put different blocks of the real twisted superfield into different group representations.  The component fields in the upper-left $N\times N$ block of $V$ are in the adjoint representation of $U(N)_1$, while those in the lower-right block are in the adjoint representation of $U(N)_2$.  The fields in the upper-right block are in an $(N,\tilde{N})$ representation.  This means that each of the $N$ columns within $\lambda_1$ is a fundamental-representation $N$-vector for the group $U(N)_1$, while at the same time, each of the $N$ rows of $\lambda_1$ is a conjugate-representation $N$-vector for the group $U(N)_2$.  The fields in the lower-left block are in the $(\tilde{N},N)$ representation.  

Even though the fields in the off-diagonal blocks are in fundamental and conjugate representations, it is convenient to expand them in terms of $t^A$ (e.g. $\lambda_1=\lambda_1^A t^A$) in order to keep notations as close as possible to those for the corresponding supersymmetric theory.

Just as for the real twisted superfield, a ``twisted'' form of the local superspace gauge function of eq \eqref{2} can be defined as follows:
\begin{equation} \label{7} 
\Lambda =\left(\begin{array}{cc} {\alpha _{1} \left(y\right)+\theta ^{2} n_{1} \left(y\right)} & {\theta \xi _{1} \left(y\right)} \\ {\theta \xi _{2} \left(y\right)} & {\alpha _{2} \left(y\right)+\theta ^{2} n_{2} \left(y\right)} \end{array}\right). 
\end{equation} 
Each of the component fields in $\Lambda$ is an $N\times N$ matrix function of $y$.  

It is straightforward to construct an action in which the real twisted superfield of eq \eqref{6} interacts with chiral superfields $Q$ and $\tilde{Q}$ in the fundamental and conjugate representations of $U(N)_1\times U(N)_2$:
\begin{equation}\label{8}
\begin{aligned}
Q =& \left(\begin{array}{c}
{\phi_1+\sqrt{2}\theta\psi_1+\theta^2f_1} \\
{\phi_2+\sqrt{2}\theta\psi_2+\theta^2f_2} 
\end{array}\right) \\
\tilde{Q} =& \left(\begin{array}{c}
{\tilde{\phi}_1+\sqrt{2}\theta\tilde{\psi}_1+\theta^2\tilde{f}_1} \\
{\tilde{\phi}_2+\sqrt{2}\theta\tilde{\psi}_2+\theta^2\tilde{f}_2} 
\end{array}\right),
\end{aligned}
\end{equation}
where each of the component fields in $Q$ and $\tilde{Q}$ is an $N$-vector function of $y$.  

Local supergauge transformations of the fields in eqs \eqref{6} and \eqref{8} are defined as follows:
\begin{equation} \label{9}
\begin{aligned}
&e^{2gV} \to e^{i\Lambda ^{\dag } } e^{2gV} e^{-i\Lambda } ,\\
&e^{-2gV^T} \to e^{-i\Lambda^{\dag T} } e^{-2gV^T} e^{i\Lambda^T  }  ,\\
&Q \to e^{i\Lambda }Q , \,\,\,\,\,\, Q^{\dag} \to Q^{\dag}e^{-i\Lambda^{\dag} }, \\
&\tilde{Q} \to e^{-i\Lambda^T }\tilde{Q}  , \,\,\,\,\,\, \tilde{Q}^{\dag} \to \tilde{Q}^{\dag} e^{i\Lambda^{\dag T} },
\end{aligned}
\end{equation}
where a $T$ superscript refers to the transpose just in group space (not in spin space).  In \cite{alternative,Twisted} it is shown that if $V$ satisfies the first expression above, then it also satisfies the second (conjugate representations are available).  There is nothing inconsistent about the above supergauge transformations: they maintain the same chirality, reality, and boson/fermion matrix structure of the fields upon which they act.  If the group chosen had been $SU(N)_1\times SU(N)_2$, on the other hand, then there would have been inconsistencies.  For example, the gauge transformation of $\lambda_1$ includes terms like $\lambda_1 \to e^{i\alpha_1} \lambda_1 e^{-i\alpha_2}+...$  Those terms are not constrained to be traceless, so the transformed $\lambda_1$ would no longer be an $SU(N)$ matrix; it would be a $U(N)$ matrix.

The part of the action involving chiral superfields is
\begin{equation} \label{10} 
S_Q  =\int d^{4} xd^{2} \theta d^{2} \bar{\theta }_{}   
\left(Q^{\dag } e^{2gV} Q
+\tilde{Q}^{\dag } e^{-2gV^T} \tilde{Q} \right).
\end{equation} 
Given eq \eqref{9}, $S_Q$ is supergauge invariant. Also, for every chiral superfield in the fundamental representation, there is a conjugate field that has opposite charge.  As a result, the above action does not produce a gauge anomaly.

The part of the action involving only the real superfield components is given by the holomorphic expression:
\begin{equation} \label{11} 
S_{V} =-{\tfrac{1}{2}}\int d^{4} xd^{2} \theta \, {\rm Tr}\left(W^{\alpha } W_{\alpha } \right) +h.c.,  
\end{equation} 
where ``Tr'' with an upper-case T denotes a trace over $2N\times 2N$ matrices, $h.c.$ denotes Hermitian conjugate, and
\begin{equation} \label{12} 
W_{\alpha } =-{\tfrac{1}{8g}} i\bar{D}_{}^{2} \left(e^{-2gV} D_{\alpha } e^{2gV} \right)   
\end{equation} 
with $D_{\alpha } =\partial _{\alpha } +i\sigma _{\alpha \dot{\alpha }}^{\mu } \bar{\theta }_{}^{\dot{\alpha }} \partial _{\mu } $.  Following standard arguments, it can be seen that $S_V$ is also supergauge invariant, so the total action is supergauge invariant.  If it can be shown that a Wess-Zumino-like gauge is accessible, then it would be possible to use gauge transformations to set all components of $V$ equal to zero other than those that are multiplied by at least one factor each of $\theta$ and $\bar{\theta}$.  It is shown below that a Wess-Zumino gauge is indeed accessible.

Starting with the form of $V$ in eq \eqref{6}, one may perform three supergauge transformations to reach a Wess-Zumino gauge.  For the first one, 
\begin{equation}\label{13}
e^{2gV} \to e^{2gV'} =e^{i\Lambda ^{\dag } } e^{2gV} e^{-i\Lambda }, 
\end{equation}
the only nonvanishing components of the gauge fields $\Lambda$ and $\Lambda^{\dag}$ are
\begin{equation}\label{14}
\alpha_m\left(y\right) =-ig C_m\left(y\right)\,\,\,\,\textrm{and}\,\,\,\,
\alpha^{\dag}_m\left(\bar{y}\right) =ig C_m\left(\bar{y}\right)	 ,			\end{equation}
where Taylor expansions can be used to rewrite these in terms of the spacetime coordinate $x$.  In the expansion of eq \eqref{13}, the contribution from terms with no factors of $\theta$ or $\bar{\theta}$ is $e^{-gC}e^{2gC}e^{-gC}=1$, where  $C$ is the block-diagonal matrix involving $C_1$ and $C_2$.  This means that when the right-hand side of eq \eqref{13} is put into the form $e^{2gV'}$ with $V'$ having the structure of eq \eqref{6}, $C'_m=0$.  The transformation of eq \eqref{13} also modifies the other component fields.  For example, $\eta'_m = e^{-gC_m}\eta_m  e^{-gC_{m' \ne m}}$.  

Two more supergauge transformations can then be imposed:
\begin{equation}\label{15}
\begin{aligned}
e^{2gV'} \to &e^{2gV''}=e^{i\Lambda ^{\prime\dag } } e^{2gV'} e^{-i\Lambda' } \,\,\,\,\textrm{with} \\
\xi'_m=&-2ig\eta'_m\,\,\,\,\textrm{and}\,\,\,\,\bar{\xi}'_m=2ig\bar{\eta}'_m ,
\end{aligned}
\end{equation}
and
\begin{equation}\label{15b}
\begin{aligned}
e^{2gV''} \to &e^{2gV'''}=e^{i\Lambda^{\prime\prime\dag } } e^{2gV''} e^{-i\Lambda'' }\,\,\,\,\textrm{with} \\
n''_m=&-2igN''_m\,\,\,\,\textrm{and}\,\,\,\,n_m^{\prime\prime\dag}=2igN_m^{\prime\prime\dag},
\end{aligned}
\end{equation}
where the components of $\Lambda'$, $\Lambda^{\prime\dag}$, $\Lambda''$, and $\Lambda^{\prime\prime\dag}$ other than those specified above vanish.  After these transformations, $V'''$ is in a Wess-Zumino-like gauge.  After dropping the triple primes in the notation, the real superfield in this Wess-Zumino gauge can be written in terms of component fields as follows:
\begin{equation} \label{16} 
V=\left(\begin{array}{cc} {-\bar{\theta }\bar{\sigma }^{\mu } A_{1\mu } \theta +{\tfrac{1}{2}} \bar{\theta }^{2} d_{1} \theta ^{2} } 
& {i\bar{\theta }\bar{\lambda}_2 \theta ^{2} -i\bar{\theta }^{2} \lambda_1 \theta } \\ 
{i\bar{\theta }\bar{\lambda}_1 \theta ^{2} -i\bar{\theta }^{2} \lambda_2\theta } & {-\bar{\theta }\bar{\sigma }^{\mu } A_{2\mu } \theta +{\tfrac{1}{2}} \bar{\theta }^{2} d_{2} \theta ^{2} } \end{array}\right) .     
\end{equation} 

In this Wess-Zumino gauge, the total action $S_Q+S_V$ is:
\begin{widetext}
\begin{equation}\label{17}
\begin{aligned}
S_Q=&\sum_m\int d^4x \left(-\phi_m^\dag D_m^\mu D_{m\mu}\phi_m-i\psi^{\dag}_m\bar{\sigma}^\mu D_{m\mu}\psi_m + f_m^\dag f_m +g\phi_m^\dag d_m \phi_m
+\left(\sqrt{2}ig\phi_m^\dag\lambda_m\psi_{m'\ne m} +h.c.\right)\right) \\
+&\sum_m\int d^4x \left(-\tilde{\phi}_m^\dag \tilde{D}_m^\mu \tilde{D}_{m\mu}\tilde{\phi}_m-i\tilde{\psi}^{\dag}_m\bar{\sigma}^\mu \tilde{D}_{m\mu}\tilde{\psi}_m + \tilde{f}_m^\dag \tilde{f}_m -g\tilde{\phi}_m^\dag d_m^T \tilde{\phi}_m
+\left(-\sqrt{2}ig\tilde{\phi}_{m'\ne m}^\dag\lambda^T_m\tilde{\psi}_m +h.c.\right)\right)
\end{aligned}
\end{equation}
\begin{equation}\label{18}
S_V=\sum_m\int d^4x {\rm{tr}}\left(-\tfrac{1}{2}F_m^{\mu\nu}F_{m\mu\nu}+d_m^2-2i\bar{\lambda}_m\bar{\sigma}^\mu\left(\partial_\mu\lambda_m-igA_{m\mu}\lambda_m+ig\lambda_m A_{m'\ne m \mu}\right)\right),
\end{equation}
\end{widetext}
where $D_{m\mu}=\partial_\mu-igA_{m \mu}$, $\tilde{D}_{m\mu}=\partial_\mu+igA^T_{m \mu}$ and ``tr'' with a lower-case t denotes a trace over $N\times N$ matrices.  This action is invariant to the following gauge transformations:
\begin{equation}\label{19}
\begin{aligned}
\phi_m&\to e^{ia_m}\phi_m \hspace{0.5cm} \psi_m\to e^{ia_m}\psi_m
\hspace{0.5cm} f_m\to e^{ia_m}f_m \\
\tilde{\phi}_m&\to e^{-ia^T_m}\tilde{\phi}_m \hspace{0.5cm} \tilde{\psi}_m\to e^{-ia^T_m}\tilde{\psi}_m
\hspace{0.5cm} \tilde{f}_m\to e^{-ia^T_m}\tilde{f}_m \\
gA_{m\mu}&\to e^{ia_m}gA_{m\mu} e^{-ia_m}-i\left(\partial_\mu e^{ia_m}\right) e^{-ia_m} \\
d_m&\to e^{ia_m}d_m e^{-ia_m}\\
\lambda_m&\to e^{ia_m}\lambda_m e^{-ia_{m'\ne m}},	
\end{aligned}
\end{equation}	 
where $a_m$ is a function of $x^\mu$.  In other words, the action in terms of component fields is a gauge theory.  Also, since fermions come in conjugate pairs, there is no gauge anomaly and the theory is renormalizable.

The theory has three different couplings that are all denoted by $g$ in eqs \eqref{17} and \eqref{18}, because they are all the same at tree level.  They are (i) the gauge coupling (multiplying gauge fields $A_m$), (ii) the ``gaugino coupling'' (multiplying $\lambda_m$ fields in eq \eqref{17}), and (iii) the ``d-term'' coupling (multiplying $d_m$ in eq \eqref{17}).  It will be shown below that for (i) and (ii) involving non-Abelian gauge fields or gauginos, all three of these couplings have the same one-loop beta function, so that they all ``run'' identically at the one-loop level.  It will also be shown below that quadratic divergences for this theory are cancelled at the two-loop level of perturbation theory.

To demonstrate this equivalence and this cancellation, the Landau gauge will be employed, resulting in a gauge propagator of $-i(g_{\mu\nu}-k_\mu k_\nu/k^2)/k^2$.  To simplify cancellations in the final parts of the analysis below, the Abelian gauge fields are redefined as:
\begin{equation}\label{20}
B_{\pm\mu}=\tfrac{1}{\sqrt{2}}\left(A^0_{1\mu}\pm A^0_{2\mu}\right).
\end{equation}

The $S_V$ part of the action can then be rewritten from eq \eqref{18} as follows:
\begin{widetext}
\begin{equation}\label{21}
\begin{aligned}
S_V&=\sum_m\int d^4x {\rm{tr}}\left(-\tfrac{1}{2}F_m^{\mu\nu}F_{m\mu\nu}+d_m^2-2i\bar{\lambda}_m\bar{\sigma}^\mu\left(\partial_\mu\lambda_m-igA_{m\mu}\lambda_m+ig\lambda_m A_{m'\ne m \mu}\right)\right) \\
&+\int d^4x \left(-\tfrac{1}{4}F_{+}^{\mu\nu}F_{+\mu\nu}-\tfrac{1}{4}F_{-}^{\mu\nu}F_{-\mu\nu} - \tfrac{1}{\sqrt{N}}gB_{-\mu}\left(\bar{\lambda}_1^A\bar{\sigma}^\mu \lambda_1^A - \bar{\lambda}_2^A\bar{\sigma}^\mu \lambda_2^A \right)\right).
\end{aligned}
\end{equation}
\end{widetext}
In eq \eqref{21}, $A_{m\mu}$ has been redefined such that it only refers to non-Abelian gauge bosons: $A_{m\mu}=A_{m\mu}^a t^a$, where lower case letters ``a,b,c'' are used to denote the $SU(N)$ indices within the $U(N)$ groups.  Upper case letters ``A,B,C'' are still used to denote the full $U(N)$ indices.  For Abelian gauge bosons, ``+'' and ``-'' are used to denote the indices of eq \eqref{20}.  For gauginos, ``0'' is used to denote the $U(1)$ indices within the $U(N)$ groups.

To show that the present theory is free of quadratic divergences to two loops, the strategy employed will be to compare the SUSY-broken theory defined by $S_Q$ and $S_V$ in eqs \eqref{17} and \eqref{21} (denoted by $\mathcal{L}$) to a supersymmetric theory that uses the real superfield of eq \eqref{1} with Abelian gauge fields rewritten using eq \eqref{20} (denoted by $\mathcal{L}_{\rm{SUSY}}$).  All one-loop diagrams of $\mathcal{L}$ and $\mathcal{L}_{\rm{SUSY}}$ will be directly compared and shown to give the same results, except for a few diagrams involving Abelian gauge fields or gauginos and a new 4-scalar vertex.  Using these one-loop results, all two-loop diagrams that generate quadratic divergences in $\mathcal{L}_{\rm{SUSY}}$ will be compared to diagrams in $\mathcal{L}$ and found to be the same.  Since $\mathcal{L}_{\rm{SUSY}}$ is free of quadratic divergences (due to supersymmetry), that same set of two-loop diagrams in $\mathcal{L}$ is also free of quadratic divergences. Some two-loop diagrams are present in $\mathcal{L}$ that are not present in $\mathcal{L}_{\rm{SUSY}}$, including ones involving the new 4-scalar vertex mentioned above and ones where Abelian gauginos and gauge fields have an interaction. It is shown that these additional diagrams all cancel.  The analysis is greatly simplified by the fact that $\mathcal{L}$ is exactly the same as $\mathcal{L}_{\rm{SUSY}}$ except for interaction terms involving gauginos. 
 
From a Feynman graph standpoint, the only differences between the supersymmetric and SUSY-broken theories are in the following diagrams:
\begin{equation}\label{22}
\begin{tabular}[t]{cc}
\includegraphics[clip=true,width=3cm]{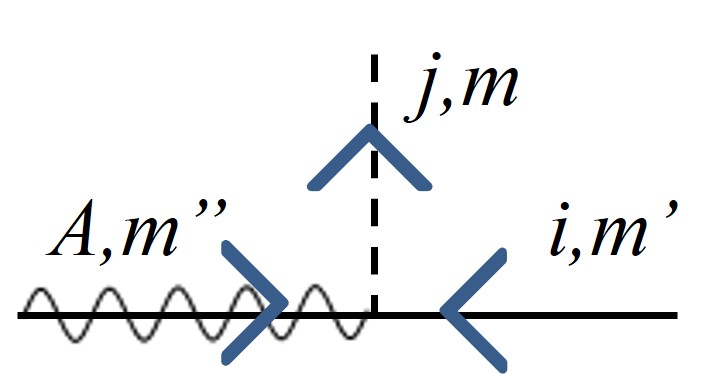}
&
{
$\begin{aligned}
&\textrm{Fundamental:} \\
\mathcal{L}_{\rm{SUSY}}\textrm{: }&-\sqrt{2}g t_{ji}^{A} \delta_{+}\\
\mathcal{L}\textrm{: }&-\sqrt{2}gt_{ji}^{A} \delta_{-} \\
&\textrm{Conjugate:} \\
\mathcal{L}_{\rm{SUSY}}\textrm{: }&\sqrt{2}g t_{ji}^{TA} \delta_{+}\\
\mathcal{L}\textrm{: }&\sqrt{2}gt_{ji}^{TA} \delta_{-}
\end{aligned}$}
\end{tabular}
\end{equation}
\begin{equation}\label{23}
\begin{tabular}[t]{cc}
\includegraphics[clip=true,width=3cm]{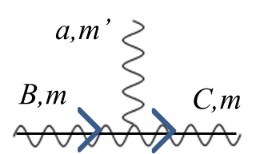}
&
{
$\begin{aligned}
\mathcal{L}_{\rm{SUSY}}\textrm{: }&g f^{aBC}\delta_{+} \\
\mathcal{L}\textrm{: }&\tfrac{1}{2}gf^{aBC} \left(\delta_{+}+\delta_{-}\right) \\
-&\tfrac{1}{2}igd^{aBC}\left(\delta_{+}-\delta_{-}\right)
\end{aligned}$}
\end{tabular}
\end{equation}
\begin{equation}\label{24}
\begin{tabular}[t]{cc}
\includegraphics[clip=true,width=3cm]{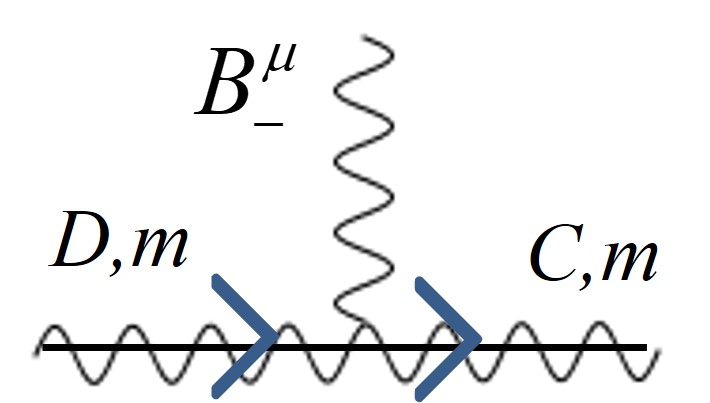}
&{$\begin{aligned}
\mathcal{L}_{\rm{SUSY}}\textrm{: }&0\\
\mathcal{L}\textrm{: }&i\tfrac{1}{\sqrt{N}}\left(-1\right)^mg\delta^{CD}
\end{aligned}$}
\end{tabular}
\end{equation}

In these diagrams (and the ones below), scalars, fermions (from chiral multiplets), gauge bosons, and gauginos are represented by dashed lines, solid lines, wavy lines, and solid+wavy lines, respectively.  The arrows point from $\phi,\lambda,\psi,\tilde{\phi},\tilde{\psi}$ and toward $\phi^\dag,\bar{\lambda},\psi^{\dag},\tilde{\phi}^{\dag},\tilde{\psi}^{\dag}$ and just the group structure is shown.  For diagram \eqref{22}, there is another diagram (not shown) that has the arrows reversed.  In these diagrams, the U(N) structure functions are defined via $f^{aBC}=-2i{\rm{tr}}\left(t^a\left[t^B,t^C\right]\right)$ and $d^{aBC}=2{\rm{tr}}\left(t^a\left\{t^B,t^C\right\}\right)$ .  Also $\delta_{+}$ means $m'=m=m''$, while $\delta_{-}$ means $m'\ne m=m''$ for fundamental fields and $m''=m'\ne m$ for conjugate fields.
	
In both theories, there are four potentially quadratically divergent one-loop diagrams:
\begin{equation}\label{25}
\begin{aligned}
&\includegraphics[clip=true]{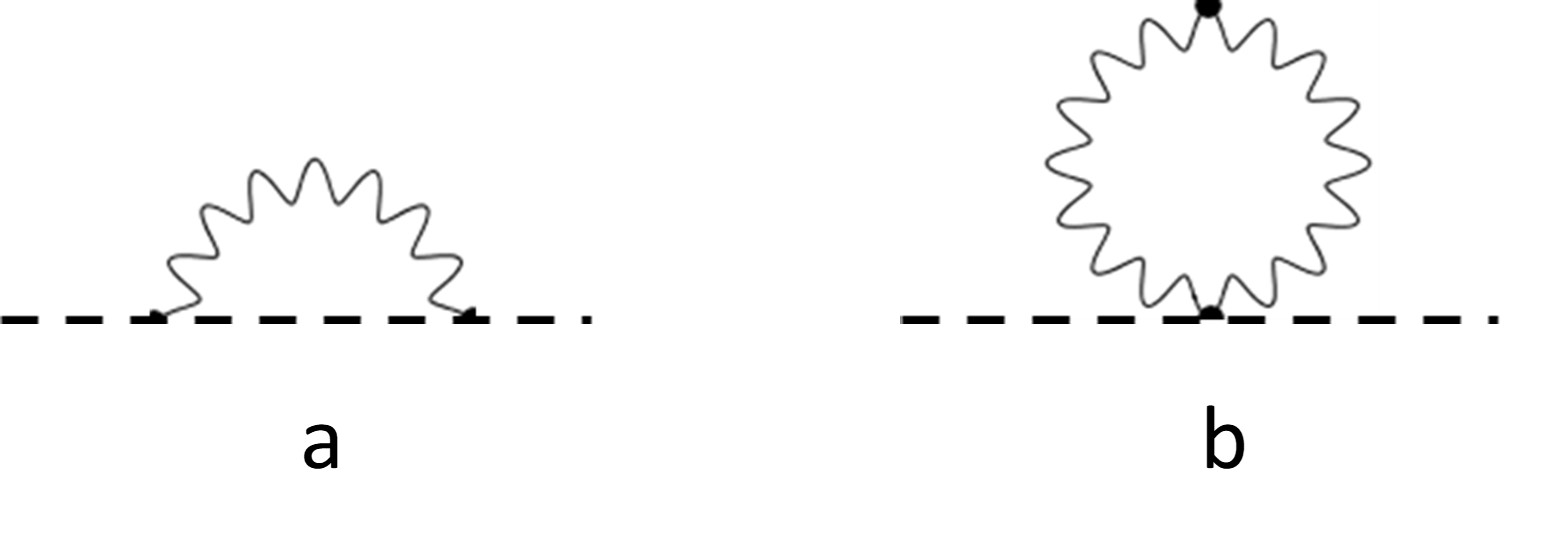} \\
&\includegraphics[clip=true]{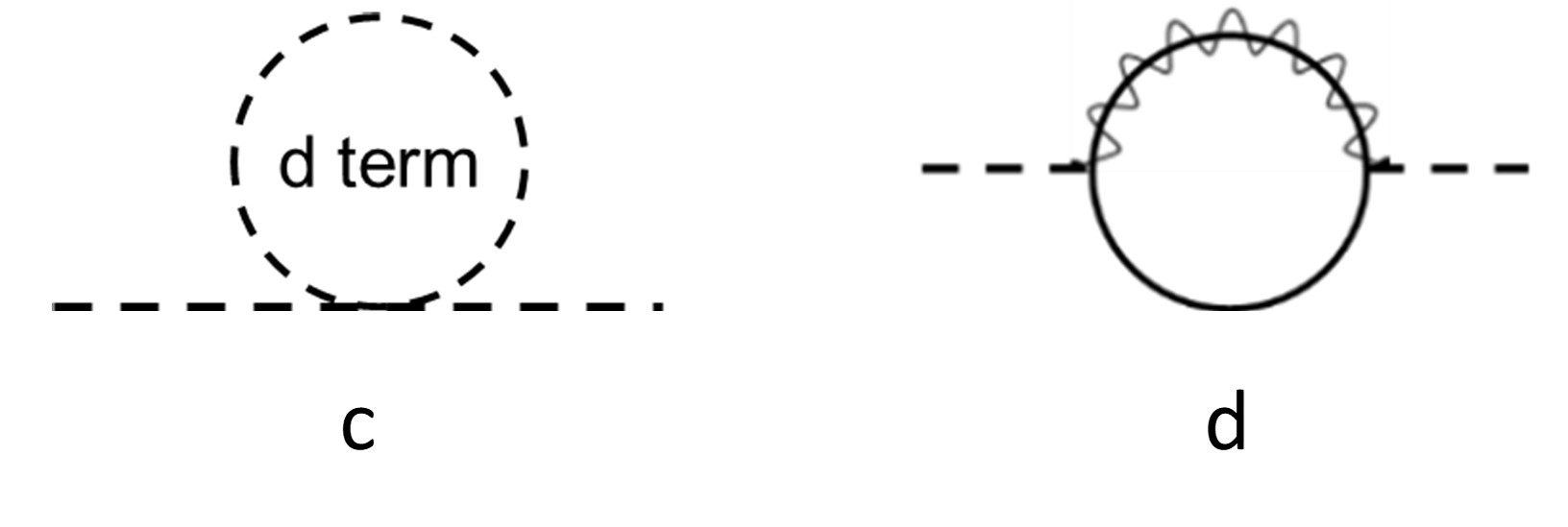}
\end{aligned}
\end{equation}
where diagram c involves the 4-scalar interaction term (the d-term) that comes from solving the equations of motion for the auxiliary $d$ field in the action.  Diagrams a, b and c generate the same results in both $\mathcal{L}$ and $\mathcal{L}_{\rm{SUSY}}$ since they do not involve gauginos.  In $\mathcal{L}_{\rm{SUSY}}$, the diagrams of \eqref{25} all cancel, so they will also cancel in $\mathcal{L}$ if diagram \eqref{25}d produces the same result in both theories.

This is indeed the case.  From eq \eqref{22}, calculation of diagram \eqref{25}d results in the following for fundamental fields:
\begin{equation}\label{26}
\begin{tabular}[t]{cc}
\includegraphics[clip=true,width=3cm]{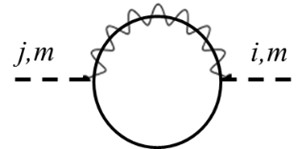}
&
{
$\begin{aligned}
&=2g^2\left(t^At^A\right)_{ij} ...\\
&\textrm{for both }\mathcal{L}\textrm{ and }\mathcal{L}_{\rm{SUSY}},
\end{aligned}$}
\end{tabular}
\end{equation}
where ``...'' stands for momentum and spin dependence which is the same in both theories.  The diagrams are also the same for conjugate fields, just replacing the above group matrices with their transposes.  Since both theories give the same contribution to all four diagrams of \eqref{25}, this calculation has verified that $\mathcal{L}$ is free of quadratic divergences at the one-loop level.  

The fact that diagram \eqref{26} gives the same results in $\mathcal{L}$ and $\mathcal{L}_{\rm{SUSY}}$ also means that the one-loop scalar wave function renormalization constants for both theories are the same.  This is because all of the other diagrams that contribute to that renormalization constant are the same in both theories due to the fact that they do not involve gauginos.

Using the same reasoning, $\mathcal{L}$ and $\mathcal{L}_{\rm{SUSY}}$ also generate the same results for the following diagram:
\begin{equation}\label{fermion}
\includegraphics[clip=true,width=4cm]{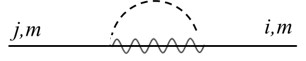}.
\end{equation}
As a result, the fermion one-loop wave function renormalization constants for both theories are the same (for both fundamental and conjugate representations).

Using eq \eqref{23} along with the identities 
\begin{equation}
\begin{aligned}
f^{acd}f^{bcd}=d^{aCD}d^{bCD}&=N\delta^{ab} \\
f^{ade}f^{bef}f^{cfd}=-f^{ade}d^{beF}d^{cFd}&=\tfrac{1}{2}Nf^{abc},
\end{aligned}
\end{equation}
it is straightforward to show that $\mathcal{L}$ and $\mathcal{L}_{\rm{SUSY}}$ generate the same results for the following one-loop diagrams:
\begin{equation}\label{28}
\begin{tabular}[t]{cc}
\includegraphics[clip=true,width=3cm]{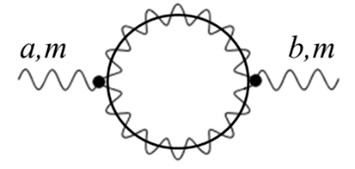}
&
\includegraphics[clip=true,width=3cm]{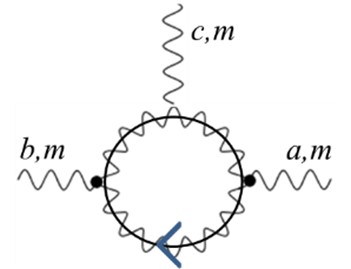}
\end{tabular}
\end{equation}
Although $\mathcal{L}$ generates terms in which some of the external gauge fields have an index $m$ while others have $m'\ne m$, all of those mixed terms exactly cancel.  The only remaining terms are those in which all of the external gauge fields have the same index $m$, and the functional form for each diagram is the same as for $\mathcal{L}_{\rm{SUSY}}$.

The fact that these diagrams produce exactly the same results in $\mathcal{L}$ and $\mathcal{L}_{\rm{SUSY}}$ means that the following are also the same in both theories: (i) the one-loop wave function renormalization constant for non-Abelian gauge bosons, (ii) the one-loop correction to the non-Abelian gauge coupling, and (iii) the one-loop beta function for the non-Abelian gauge coupling.  Also, since both $\mathcal{L}$ and $\mathcal{L}_{\rm{SUSY}}$ are gauge theories, the non-Abelian coupling constant of a 3-gauge-boson vertex is the same as that of the four-point vertex in diagram \eqref{25}b.  As a result, all 2-loop diagrams for $\mathcal{L}$ that involve adding another loop to diagram \eqref{25}b when the gauge boson in the loop is non-Abelian give the same results as the corresponding 2-loop diagrams in $\mathcal{L}_{\rm{SUSY}}$.

It can also be shown that $\mathcal{L}$ and $\mathcal{L}_{\rm{SUSY}}$ produce the same result for the following one-loop diagrams:
\begin{equation}\label{29}
\begin{aligned}
&\begin{tabular}[t]{cc}
&\includegraphics[clip=true,width=4cm]{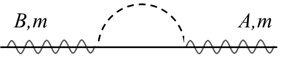} 
\end{tabular}\\
&\begin{tabular}[t]{cc}
\includegraphics[clip=true,width=3cm]{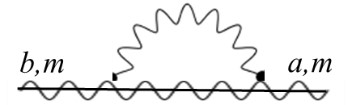} .
&
\hspace{3cm}
\end{tabular}
\end{aligned}
\end{equation}
As a result, the one-loop wave function renormalization constant for non-Abelian gauginos is the same in both theories.  For Abelian gauginos, the results are different between $\mathcal{L}$ and $\mathcal{L}_{\rm{SUSY}}$, but in $\mathcal{L}$, Abelian gauginos have the same one-loop wave function renormalization constant as non-Abelian gauginos.

For the following diagram,
\begin{equation}\label{30}
\begin{tabular}[t]{cc}
\includegraphics[clip=true,width=3cm]{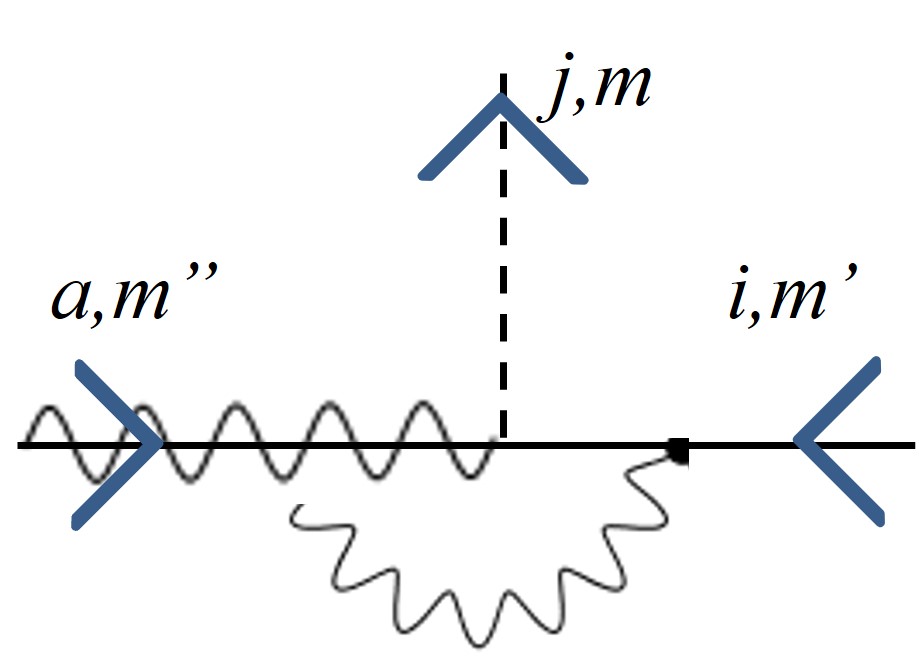}
&
\hspace{3cm}
\end{tabular}
\end{equation}
the magnitude of the correction is the same in both theories; the only difference is that for $\mathcal{L}_{\rm{SUSY}}$, $m'=m=m''$, while for $\mathcal{L}$ with fundamental fields, $m'\ne m=m''$, and for $\mathcal{L}$ with conjugate fields, $m''=m'\ne m$, matching their tree-level behavior in eq \eqref{22}. Since the one-loop wave function renormalization constants for scalars, fermions, and non-Abelian gauginos is also the same in both theories, the one-loop beta function for this coupling is the same in $\mathcal{L}$ and $\mathcal{L}_{\rm{SUSY}}$.  It also means that all 2-loop diagrams for $\mathcal{L}$ that involve adding another loop to diagram \eqref{25}d when the gauginos in the loop are non-Abelian give the same result as the corresponding 2-loop diagrams in $\mathcal{L}_{\rm{SUSY}}$.

The vertex corrections of \eqref{30} using Abelian gauginos (instead of non-Abelian) are different between $\mathcal{L}$ and $\mathcal{L}_{\rm{SUSY}}$.  But in $\mathcal{L}$ the vertex with Abelian gauginos has the same one-loop correction as the vertex for non-Abelian gauginos.  As a result, the one-loop beta function for the gaugino coupling in $\mathcal{L}$ is the same when the gaugino is Abelian as when it is non-Abelian.

It is next interesting to consider the one-loop correction to the d-term vertex:
\begin{equation}\label{27}
\begin{tabular}[t]{cc}
\includegraphics[clip=true,width=3cm]{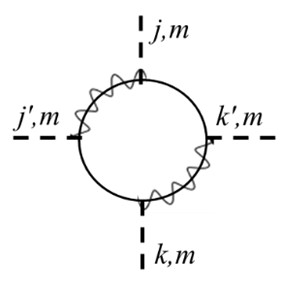} .
&
\hspace{3cm}
\end{tabular}
\end{equation}
For this diagram, the two theories generate different results. $\mathcal{L}$ generates the same one-loop correction to the d-term vertex as $\mathcal{L}_{\rm{SUSY}}$, but it also generates a new vertex proportional to:
\begin{equation}\label{dtermloop}
g^4(\phi_1^\dag t^A\phi_1-\phi_2^\dag t^A\phi_2)(\tilde{\phi}_1^\dag t^A\tilde{\phi}_1-\tilde{\phi}_2^\dag t^A\tilde{\phi}_2).
\end{equation}
Since the one-loop correction to the d-term vertex and the one-loop scalar wave function renormalization constant are the same in both theories, the one-loop beta function for the d-term coupling is the same in $\mathcal{L}$ and $\mathcal{L}_{\rm{SUSY}}$.  

Using the new vertex of eq \eqref{dtermloop}, a 2-loop correction to the $\phi_1^\dag\phi_1$ 2-point function can be found by contracting over $\tilde{\phi}_1^\dag\tilde{\phi}_1$ or over $\tilde{\phi}_2^\dag\tilde{\phi}_2$.  But these two contributions exactly cancel.  By symmetry, the new vertex does not generate any 2-loop contribution to scalar 2-point functions.  Combining this with the fact that the d-term vertex correction is the same in both theories means that all 2-loop corrections to the scalar propagator that involve adding another loop to diagram \eqref{25}c are the same in $\mathcal{L}$ and $\mathcal{L}_{\rm{SUSY}}$.

In the Landau gauge, none of the 2-loop diagrams that involve adding another loop to diagram \eqref{25}a are quadratically divergent.  Combining this fact with the results above means that all of the quadratically divergent two-loop diagrams in $\mathcal{L}_{\rm{SUSY}}$ are also present in $\mathcal{L}$ and have the same values.  Since those diagrams cancel in $\mathcal{L}_{\rm{SUSY}}$, they also cancel in $\mathcal{L}$.  But $\mathcal{L}$ has some additional quadratically divergent two-loop diagrams that are not present in $\mathcal{L}_{\rm{SUSY}}$: ones that involve either the vertex of eq \eqref{24} or a vertex involving an Abelian gaugino and a gauge boson.  These additional diagrams are found by (i) adding a gaugino loop to diagram \eqref{25}b when the gauge boson in the diagram is $B_{-\mu}$, or (ii) adding a loop to diagram \eqref{25}d that involves a vertex with an Abelian gaugino and a gauge boson. These additional diagrams must be shown to cancel among themselves in $\mathcal{L}$.

In other words in $\mathcal{L}$, the following three diagrams must cancel:
\begin{equation}\label{31}
\begin{tabular}[t]{ccc}
a\,\,\includegraphics[clip=true,width=2.5cm]{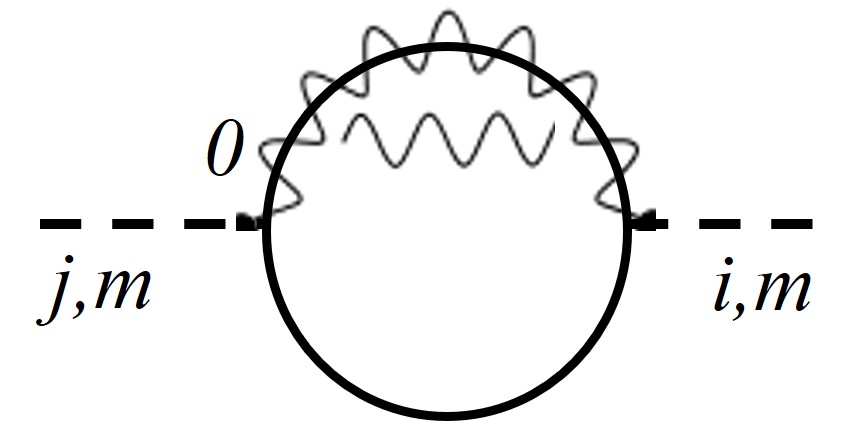}
&
b\,\,\includegraphics[clip=true,width=2.5cm]{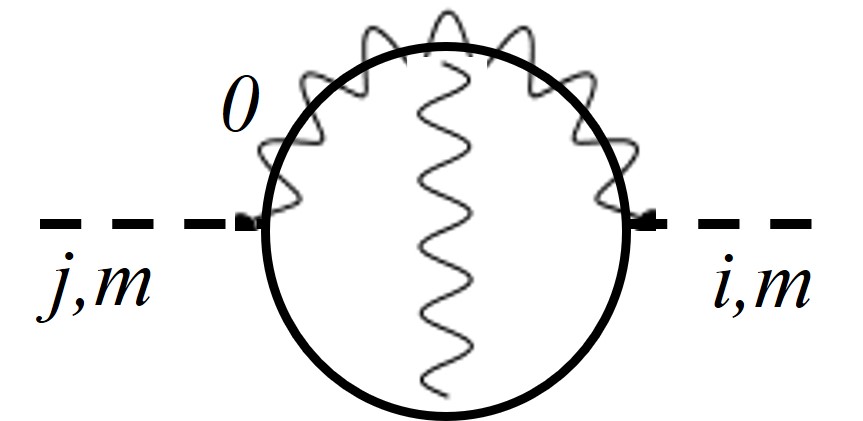}
&
c\,\,\includegraphics[clip=true,width=2.5cm]{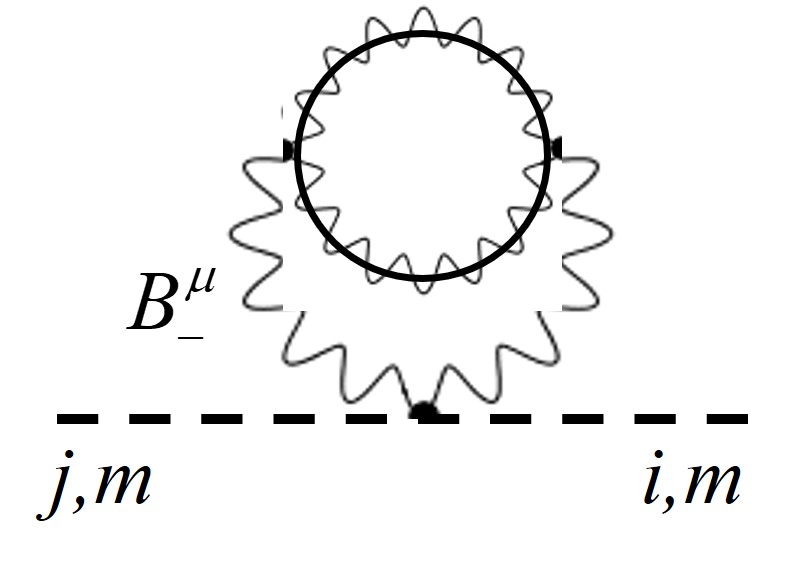}
\end{tabular}
\end{equation}	
where in diagrams a and b, $0$ denotes an Abelian gaugino, and in c, the gauge boson is $B_{-}^\mu$.  

For zero external momenta and in the Landau gauge, these diagrams are proportional to
\begin{equation}\label{diags}
i\delta_{ji}\int\frac{d^4kd^4q}{(2\pi)^8}\left(\frac{k^2+q^2}{k^4q^4}-3\frac{k^4+q^4}{k^4(k+q)^2q^4}+2\frac{k^2+q^2}{k^4q^4}\right),
\end{equation}
where the first, second and third terms are for diagrams a, b, and c. For diagram c, the gaugino loop has contributions from both $\lambda_1$ and $\lambda_2$. The spin dependence of diagrams a and c generates factors of $k\cdot q$ in their numerators. These factors can be rewritten as $2k\cdot q=(k+q)^2-k^2-q^2$ and the $(k+q)^2$ part cancels the same factor in the denominator.  For all three diagrams, $(k^2-q^2)^2$ vanishes in the numerator for a denominator that is symmetric in $k\leftrightarrow q$. 

By shifting the integrals $k+q\to k'$ in the diagram b part of eq \eqref{diags}, it can be seen that the contribution from diagram b exactly cancels the contributions from diagrams a and c.  The cancellation of these three diagrams completes the proof that the theory is free of 2-loop quadratic divergences.

In addition, it was shown above that three of the couplings in $\mathcal{L}$ (non-Abelian gauge, non-Abelian gaugino, and d-term) have one-loop beta functions that are the same as those in $\mathcal{L}_{\rm{SUSY}}$.  Since those beta functions are all equal in $\mathcal{L}_{\rm{SUSY}}$, they are also all equal in $\mathcal{L}$. 

It was mentioned above that this cancellation takes place at a unification scale where all Abelian and non-Abelian coupling constants are the same.  The reason for that limitation is because it has not been proven that there is a symmetry that keeps beta functions for Abelian couplings the same as those for non-Abelian couplings.  However, it was noted above that the one-loop beta function for the gaugino coupling is the same for both Abelian and non-Abelian gauginos.  

This paper has shown that it is possible to create a theory with broken supersymmetry that is free of quadratic divergences to at least two loops and where “superpartners” are in different representations of the gauge group.  In \cite{alternative,Twisted}, it was shown that a theory similar to this one can reproduce the symmetries, charges and particles of the Standard Model with fewer unobserved particles than in supersymmetric theories, since in that theory, the currently observed particles are “superpartners” with each other in “twisted superfields”.

It is possible that if the model presented in this paper is indeed truly ``soft'' (free of quadratic divergences to all orders), it could be the low-energy effective theory of a more general supersymmetric theory that is broken by one of the known soft SUSY-breaking mechanisms.


\begin{thebibliography}{13}%
\makeatletter
\providecommand \@ifxundefined [1]{%
 \@ifx{#1\undefined}
}%
\providecommand \@ifnum [1]{%
 \ifnum #1\expandafter \@firstoftwo
 \else \expandafter \@secondoftwo
 \fi
}%
\providecommand \@ifx [1]{%
 \ifx #1\expandafter \@firstoftwo
 \else \expandafter \@secondoftwo
 \fi
}%
\providecommand \natexlab [1]{#1}%
\providecommand \enquote  [1]{``#1''}%
\providecommand \bibnamefont  [1]{#1}%
\providecommand \bibfnamefont [1]{#1}%
\providecommand \citenamefont [1]{#1}%
\providecommand \href@noop [0]{\@secondoftwo}%
\providecommand \href [0]{\begingroup \@sanitize@url \@href}%
\providecommand \@href[1]{\@@startlink{#1}\@@href}%
\providecommand \@@href[1]{\endgroup#1\@@endlink}%
\providecommand \@sanitize@url [0]{\catcode `\\12\catcode `\$12\catcode
  `\&12\catcode `\#12\catcode `\^12\catcode `\_12\catcode `\%12\relax}%
\providecommand \@@startlink[1]{}%
\providecommand \@@endlink[0]{}%
\providecommand \url  [0]{\begingroup\@sanitize@url \@url }%
\providecommand \@url [1]{\endgroup\@href {#1}{\urlprefix }}%
\providecommand \urlprefix  [0]{URL }%
\providecommand \Eprint [0]{\href }%
\providecommand \doibase [0]{https://doi.org/}%
\providecommand \selectlanguage [0]{\@gobble}%
\providecommand \bibinfo  [0]{\@secondoftwo}%
\providecommand \bibfield  [0]{\@secondoftwo}%
\providecommand \translation [1]{[#1]}%
\providecommand \BibitemOpen [0]{}%
\providecommand \bibitemStop [0]{}%
\providecommand \bibitemNoStop [0]{.\EOS\space}%
\providecommand \EOS [0]{\spacefactor3000\relax}%
\providecommand \BibitemShut  [1]{\csname bibitem#1\endcsname}%
\let\auto@bib@innerbib\@empty
\bibitem [{\citenamefont {Girardello}\ and\ \citenamefont
  {Grisaru}(1982)}]{Grisaru-soft}%
  \BibitemOpen
  \bibfield  {author} {\bibinfo {author} {\bibfnamefont {L.}~\bibnamefont
  {Girardello}}\ and\ \bibinfo {author} {\bibfnamefont {M.}~\bibnamefont
  {Grisaru}},\ }\bibfield  {title} {\bibinfo {title} {Soft breaking of
  supersymmetry},\ }\href@noop {} {\bibfield  {journal} {\bibinfo  {journal}
  {Nuclear Physics B}\ }\textbf {\bibinfo {volume} {194}} (\bibinfo {year}
  {1982})}\BibitemShut {NoStop}%
\bibitem [{\citenamefont {Martin}(2000)}]{Martin-soft}%
  \BibitemOpen
  \bibfield  {author} {\bibinfo {author} {\bibfnamefont {S.}~\bibnamefont
  {Martin}},\ }\bibfield  {title} {\bibinfo {title} {Dimensionless
  supersymmetry breaking couplings, flat directions, and the origin of
  intermediate mass scales},\ }\href@noop {} {\bibfield  {journal} {\bibinfo
  {journal} {Phys. Rev. D}\ }\textbf {\bibinfo {volume} {61}} (\bibinfo {year}
  {2000})},\ \Eprint {https://arxiv.org/abs/hep-ph/9907550} {hep-ph/9907550}
  \BibitemShut {NoStop}%
\bibitem [{\citenamefont {Gunion}\ and\ \citenamefont
  {Haber}(1986)}]{MSSM-1986}%
  \BibitemOpen
  \bibfield  {author} {\bibinfo {author} {\bibfnamefont {J.}~\bibnamefont
  {Gunion}}\ and\ \bibinfo {author} {\bibfnamefont {H.}~\bibnamefont {Haber}},\
  }\bibfield  {title} {\bibinfo {title} {{Higgs Bosons in Supersymmetric
  Models}},\ }\href@noop {} {\bibfield  {journal} {\bibinfo  {journal} {Nucl.
  Phys. B}\ }\textbf {\bibinfo {volume} {272}},\ \bibinfo {pages} {1} (\bibinfo
  {year} {1986})}\BibitemShut {NoStop}%
\bibitem [{\citenamefont {Haber}\ and\ \citenamefont {Kane}(1985)}]{MSSM-1985}%
  \BibitemOpen
  \bibfield  {author} {\bibinfo {author} {\bibfnamefont {H.}~\bibnamefont
  {Haber}}\ and\ \bibinfo {author} {\bibfnamefont {G.}~\bibnamefont {Kane}},\
  }\bibfield  {title} {\bibinfo {title} {{The Search for Supersymmetry: Probing
  Physics Beyond the Standard Model}},\ }\href@noop {} {\bibfield  {journal}
  {\bibinfo  {journal} {Phys. Rept.}\ }\textbf {\bibinfo {volume} {117}},\
  \bibinfo {pages} {75} (\bibinfo {year} {1985})}\BibitemShut {NoStop}%
\bibitem [{\citenamefont {Djouadi}(2008)}]{MSSM-2008}%
  \BibitemOpen
  \bibfield  {author} {\bibinfo {author} {\bibfnamefont {A.}~\bibnamefont
  {Djouadi}},\ }\bibfield  {title} {\bibinfo {title} {{The Anatomy of
  Electro-Weak Symmetry Breaking. II: The Higgs bosons in the Minimal
  Supersymmetric Model}},\ }\href@noop {} {\bibfield  {journal} {\bibinfo
  {journal} {Phys. Rept.}\ }\textbf {\bibinfo {volume} {459}},\ \bibinfo
  {pages} {1} (\bibinfo {year} {2008})}\BibitemShut {NoStop}%
\bibitem [{\citenamefont {Haag}\ \emph {et~al.}(1975)\citenamefont {Haag},
  \citenamefont {Lopuszanski},\ and\ \citenamefont {Sohnius}}]{HLS}%
  \BibitemOpen
  \bibfield  {author} {\bibinfo {author} {\bibfnamefont {R.}~\bibnamefont
  {Haag}}, \bibinfo {author} {\bibfnamefont {J.}~\bibnamefont {Lopuszanski}},\
  and\ \bibinfo {author} {\bibfnamefont {M.}~\bibnamefont {Sohnius}},\
  }\bibfield  {title} {\bibinfo {title} {{All Possible Generators of
  Supersymmetries of the S-Matrix}},\ }\href@noop {} {\bibfield  {journal}
  {\bibinfo  {journal} {Nuclear Physics B}\ }\textbf {\bibinfo {volume} {88}},\
  \bibinfo {pages} {257} (\bibinfo {year} {1975})}\BibitemShut {NoStop}%
\bibitem [{\citenamefont {Argyres}(2001)}]{SQCD-Argyres}%
  \BibitemOpen
  \bibfield  {author} {\bibinfo {author} {\bibfnamefont {P.}~\bibnamefont
  {Argyres}},\ }\href
  {http://homepages.uc.edu/~argyrepc/cu661-gr-SUSY/susy2001.pdf} {\bibinfo
  {title} {Intro to global supersymmetry}} (\bibinfo {year} {2001}),\ \bibinfo
  {note} {{Cornell University Course}}\BibitemShut {NoStop}%
\bibitem [{\citenamefont {Argurio}(2017)}]{rargurio}%
  \BibitemOpen
  \bibfield  {author} {\bibinfo {author} {\bibfnamefont {R.}~\bibnamefont
  {Argurio}},\ }\href {http://homepages.ulb.ac.be/~rargurio/susycourse.pdf}
  {\bibinfo {title} {Phys-f-417 supersymmetry course}} (\bibinfo {year}
  {2017}),\ \bibinfo {note} {{Lecture notes from Universite Libre de
  Bruxelles}}\BibitemShut {NoStop}%
\bibitem [{\citenamefont {Martin}(2016)}]{SUSY-Martin}%
  \BibitemOpen
  \bibfield  {author} {\bibinfo {author} {\bibfnamefont {S.}~\bibnamefont
  {Martin}},\ }\href@noop {} {\bibinfo {title} {{A Supersymmetry Primer}}}
  (\bibinfo {year} {2016}),\ \Eprint {https://arxiv.org/abs/hep-ph/9709356v7}
  {hep-ph/9709356v7} \BibitemShut {NoStop}%
\bibitem [{\citenamefont {Haber}\ and\ \citenamefont
  {Haskins}(2018)}]{SUSY-Haber}%
  \BibitemOpen
  \bibfield  {author} {\bibinfo {author} {\bibfnamefont {H.}~\bibnamefont
  {Haber}}\ and\ \bibinfo {author} {\bibfnamefont {L.}~\bibnamefont
  {Haskins}},\ }\href@noop {} {\bibinfo {title} {{Supersymmetric Theory and
  Models}}} (\bibinfo {year} {2018}),\ \Eprint
  {https://arxiv.org/abs/hep-ph/1712.05926v4} {hep-ph/1712.05926v4}
  \BibitemShut {NoStop}%
\bibitem [{\citenamefont {Binetruy}(2006)}]{Binetruy}%
  \BibitemOpen
  \bibfield  {author} {\bibinfo {author} {\bibfnamefont {P.}~\bibnamefont
  {Binetruy}},\ }\href@noop {} {\emph {\bibinfo {title} {Supersymmetry: Theory,
  Experiment, and Cosmology}}}\ (\bibinfo  {publisher} {Oxford University
  Press},\ \bibinfo {year} {2006})\BibitemShut {NoStop}%
\bibitem [{\citenamefont {Chapman}(2022)}]{alternative}%
  \BibitemOpen
  \bibfield  {author} {\bibinfo {author} {\bibfnamefont {S.}~\bibnamefont
  {Chapman}},\ }\bibfield  {title} {\bibinfo {title} {{An Alternative to the
  Standard Model}},\ }\href {https://doi.org/10.1007/s40509-021-00268-4}
  {\bibfield  {journal} {\bibinfo  {journal} {Quant. Studies: Math and
  Foundations}\ }\textbf {\bibinfo {volume} {9}},\ \bibinfo {pages} {235}
  (\bibinfo {year} {2022})}\BibitemShut {NoStop}%
\bibitem [{\citenamefont {Chapman}(2023)}]{Twisted}%
  \BibitemOpen
  \bibfield  {author} {\bibinfo {author} {\bibfnamefont {S.}~\bibnamefont
  {Chapman}},\ }\href@noop {} {\bibinfo {title} {Twisted superfields}}
  (\bibinfo {year} {2023}),\ \bibinfo {note} {an update to ``An alternative to
  the Standard Model''},\ \Eprint {https://arxiv.org/abs/2112.04469}
  {2112.04469} \BibitemShut {NoStop}%
\end{thebibliography}
\end{document}